\documentclass[nofootinbib,prd,showpacs]{revtex4-1}
\usepackage{amsmath}
\usepackage{amsfonts}
\usepackage{amssymb}
\usepackage{graphicx}
\graphicspath{ {Fig/} }
\usepackage[usenames,dvipsnames]{xcolor}
\usepackage{hyperref}   
\hypersetup{colorlinks=true, 
citecolor=MidnightBlue, linkcolor=MidnightBlue, urlcolor=Cyan}
\usepackage{tikz}
\definecolor{purple}{rgb}{1,0,1}
\definecolor{lime}{HTML}{A6CE39} 

\newcommand{\orcidicon}{%
	\begin{tikzpicture}
	\draw[lime, fill=lime] (0,0) 
		circle [radius=0.16] 
		node[white] {{\fontfamily{qag}\selectfont \tiny ID}};
	\draw[white, fill=white] (-0.0625,0.095) 
		circle [radius=0.007];
	\end{tikzpicture}	\hspace{-2mm}
}
\newcommand\orcidMiguel{{\href{https://orcid.org/0000-0003-3367-9868}{\orcidicon}}}
\newcommand\orcidFrancisco{{\href{https://orcid.org/0000-0002-9388-8373}{\orcidicon}}}
\begin{document}
\title{Warp drive basics}
\author{Miguel Alcubierre\orcidMiguel\!\!}
\email{malcubi@nucleares.unam.mx}\affiliation{Instituto de Ciencias Nucleares, Universidad Nacional Aut\'{o}noma de M\'{e}xico, Circuito Exterior C.U., A.P. 70-543, M\'{e}xico D.F. 04510, 
M\'{e}xico}
\author{Francisco S. N. Lobo\orcidFrancisco\!\!}
\email{fslobo@fc.ul.pt}
\affiliation{Instituto de Astrof\'{i}sica e Ci\^{e}ncias do Espa\c{c}o, Faculdade de Ci\^encias da Universidade de Lisboa, Edif\'{i}cio C8, Campo Grande, P-1749-016, Lisbon, Portugal}
\date{\LaTeX-ed \today}
\begin{abstract}
``Warp drive'' spacetimes and wormhole geometries are useful as ``gedanken-experiments'' that force us to confront the foundations of general relativity, and among other issues, to precisely formulate the notion of ``superluminal'' travel and communication. Here we will consider the basic definition and properties of warp drive spacetimes. In particular, we will discuss the violation of the energy conditions associated with these spacetimes, as well as some other interesting properties such as the appearance of horizons for the superluminal case, and the possibility of using a warp drive to create closed timelike curves. Furthermore, due to the horizon problem, an observer in a spaceship cannot create nor control on demand a warp bubble. To contour this difficulty, we discuss a metric introduced by Krasnikov, which also possesses the interesting property in that the time for a round trip, as measured by clocks at the starting point, can be made arbitrarily short. 
\end{abstract}
\maketitle
\def\HMS{{\scriptscriptstyle{\rm HMS}}}
\bigskip
\hrule
\tableofcontents
\bigskip
\hrule
\parindent0pt
\parskip7pt
\vspace{-10pt}

\section{Introduction}

Recently much interest has been revived in superluminal travel, due to the research in wormhole geometries \cite{Morris,VisserAL} and superluminal warp drive spacetimes \cite{Alcubierre}. However, despite the use of the term superluminal, it is not possible to locally achieve faster than light travel. In fact, the point to note is that one can make a round trip, between two points separated by a distance $D$, in an arbitrarily short time as measured by an observer that remained at rest at the starting point, by varying one's speed or by changing the distance one is to cover. It is a highly nontrivial issue to provide a general global definition of superluminal travel \cite{VB,VBL}, but it has been shown that the spacetimes that permit ``effective'' superluminal travel generically suffer from the several severe drawbacks. In particular, superluminal effects are associated with the presence of \emph{exotic} matter, that is, matter that violates the null energy condition (NEC).

In fact, it has been shown that superluminal spacetimes violate all the known energy conditions and, in particular, it was shown that negative energy densities and superluminal travel are intimately related~\cite{Olum}. Although it is thought that most classical forms of matter obey the energy conditions, they are violated by certain quantum fields~\cite{VisserEC}. Additionally, specific classical systems (such as non-minimally coupled scalar fields) have been found that violate the null and the weak energy conditions~\cite{barcelovisser1,barcelovisserPLB99}. It is also interesting to note that recent observations in cosmology, such as the late-time cosmic speed-up ~\cite{Riess}, strongly suggest that the cosmological fluid violates the strong energy condition (SEC), and provides tantalizing hints that the NEC is violated in a classical regime~\cite{Riess,jerk,rip}.

In addition to womrhole geometries {\cite{Morris,VisserAL}}, other spacetimes that allow superluminal travel are the Alcubierre warp drive \cite{Alcubierre} and the {\it Krasnikov tube} \cite{Krasnikov,Everett}, which will be presented in detail below. Indeed, it was shown theoretically shown that the Alcubierre warp drive entails the possibility to attain arbitrarily large velocities \cite{Alcubierre}, within the framework of general relativity. As will be demonstrated below, a warp bubble is driven by a local expansion of space behind the bubble, and an opposite contraction ahead of it. However, by introducing a slightly more complicated metric, Nat\'{a}rio~\cite{Natario} dispensed with the need for expansion of the volume elements. In the Nat\'{a}rio warp drive the expansion (contraction) of the distances along the direction of motion is compensated by a contraction (expansion) of area elements in the perpendicular direction, so that the volume elements are preserved.  Thus, the essential property of the warp drive is revealed to be the change in distances along the direction of motion, and not the expansion/contraction of space. Thus, the Nat\'{a}rio version of the warp drive can be thought of as a bubble sliding through space.

However, an interesting aspect of the Alcubierre warp drive is that an observer on a spaceship, within the warp bubble, cannot create nor control on demand a superluminal Alcubierre bubble surrounding the ship \cite{Krasnikov}. This is due to the fact points on the outside front edge of the bubble are always spacelike separated from the centre of the bubble. Note that, In principle, causality considerations do not prevent the crew of a spaceship from altering the metric along the path of their outbound trip, by their own actions, in order to complete a round trip from the Earth to a distant star and back in an arbitrarily short time, as measured by clocks on the Earth. To this effect, an interesting solution was introduced by Krasnikov, that consists of a two-dimensional metric with the property that although the time for a one-way trip to a distant destination cannot be shortened, the time for a round trip, as measured by clocks at the starting point (e.g. Earth), can be made arbitrarily short. Soon after, Everett and Roman generalized the Krasnikov two-dimensional analysis to four dimensions, denoting the solution as the {\it Krasnikov tube}~\cite{Everett}. Interesting features were analyzed, such as the effective superluminal nature of the solution, the energy condition violations, the appearance of closed timelike curves and the application of the Quantum Inequality (QI) deduced by Ford and Roman~\cite{Ford:1994bjAL}.

Using the QI in the context of warp drive spacetimes, it was soon verified that enormous amounts of energy are needed to sustain superluminal warp drive spacetimes~\cite{Ford:1995wg,PfenningF}. However, one should note the fact that the quantum inequalities might not necessarily be fundamental, and anyway they are violated in the Casimir effect. To reduce the enormous amounts of exotic matter needed in the superluminal warp drive, van den Broeck proposed a slight modification of the Alcubierre metric that considerably improves the conditions of the solution~\cite{VanDenBroeck:1999sn}. It was also shown that using the QI enormous quantities of negative energy densities are needed to support the superluminal Krasnikov tube~\cite{Everett}. This problem was surpassed by Gravel and Plante~\cite{GravelPlante,Gravel}, who in similar manner to the van den Broeck analysis, showed that it is theoretically possible to lower significantly the mass of the Krasnikov tube.  

However, applying the linearized approach to the warp drive spacetime \cite{LV-CQG}, where no {\it a priori} assumptions as to the ultimate source of the energy condition violations are required, the QI are not used nor needed. This means that the linearized restrictions derived on warp drive spacetimes are more generic than those derived using the quantum inequalities, where the restrictions derived in \cite{LV-CQG} hold regardless of whether the warp drive is assumed to be classical or quantum in its operation. It was not meant to suggest that such a {\it reaction-less drive} is achievable with current technology, as indeed extremely stringent conditions on the warp bubble were obtained, in the weak-field limit. These conditions are so stringent that it appears unlikely that the warp drive will ever prove technologically useful.

This chapter is organized in the following manner: In section \ref{ALsec:2}, we present the basics of the warp drive spacetime, showing the explicit violations of the energy conditions, and a brief application of the QI. In section \ref{ALsec:3}, using linearized theory, we show that significant constraints in the weak-field regime arise, so that the analysis implies additional difficulties for developing a ``strong field'' warp drive. In section \ref{ALsec:4}, we consider further interesting aspects of the warp drive spacetime, such as the ``horizon problem'', in which an observer on a spaceship cannot create nor control on demand a superluminal Alcubierre bubble. In section \ref{ALsec:5}, we consider the the physical properties of the Krasnikov tube, that consists of a metric in which the time for a round trip, as measured by clocks at the starting point, can be made
arbitrarily short. In section \ref{ALsec:6}, we consider the possibility of closed timelike curves in the superluminal warp drive and the Krasnikov tube, and in section \ref{ALsec:conclusion}, we conclude.

\section{Warp drive spacetime}\label{ALsec:2}

Alcubierre proved that it is, in principle, possible to warp spacetime in a small {\it bubble-like} region, within the framework of general relativity, in a manner that the bubble may attain arbitrarily large velocities. The enormous speed of separation arises from the expansion of spacetime itself, analogously to the inflationary phase of the early universe. More specifically, the hyper-fast travel is induced by creating a local distortion of spacetime, producing a contraction ahead of the bubble, and an opposite expansion behind ahead it.

\subsection{Alcubierre warp drive}\label{Alcubierrewarp}

In cartesian coordinates, the Alcubierre warp drive spacetime metric is given by (with $G=c=1$)
\begin{equation}
d s^2=-d t^2+d x^2+d y^2+\left[d z-v(t)\;f(x,y,z-z_0(t))\; d
t\right]^2 \label{Cartesianwarpmetric}\,,
\end{equation}
where $v(t)=dz_0(t)/dt$ is the velocity of the warp bubble, moving along the positive $z$-axis. The form function $f(x,y,z)$ possesses the general features of
having the value $f=0$ in the exterior and $f=1$ in the interior
of the bubble. The general class of form functions, $f(x,y,z)$,
chosen by Alcubierre was spherically symmetric: $f(r)$ with
$r=\sqrt{x^2+y^2+z^2}$. Then $f(x,y,z-z_0(t)) = f(r(t))$ with
$r(t)=\left\{[(z-z_{0}(t)]^2+x^2+y^2\right\}^{1/2}$.

We consider the specific case given by
\begin{equation}
f(r)=\frac{\tanh\left[\sigma(r+R)\right]
-\tanh\left[\sigma(r-R)\right]}{2\tanh(\sigma R)}\,,
\label{E:form}
\end{equation}
where $R>0$ and $\sigma>0$ are two arbitrary parameters. $R$ is
the radius of the warp-bubble, and $\sigma$ can be interpreted
as being inversely proportional to the bubble wall thickness. If
$\sigma$ is sufficiently large, the form function rapidly approaches a {\it top
hat} function, i.e., $f(r)=1$ if $r\in[0,R]$, and $f(r)=0$ if $r\in(R,\infty)$, for $\sigma \rightarrow \infty$.

It can be shown that observers with the four velocity
\begin{equation}
U^{\mu}=\left(1,0,0,vf\right), 
\end{equation}
and $U_{\mu}=\left(-1,0,0,0\right)$ move along geodesics, as their $4$-acceleration is zero, \emph{i.e.},
$a^{\mu} = U^{\nu}\; U^{\mu}{}_{;\nu}=0$. These observers are usually called ``Eulerian observers'' in the 3+1 formalism, as they move along the normal directions to the spatial slices \cite{Alcubierre}. The spaceship, which in the original formulation is treated as a test particle which moves along the curve $z=z_0(t)$, can easily be seen to always move along a timelike curve, regardless of the value
of $v(t)$. One can also verify that the proper time along this curve
equals the coordinate time, by simply substituting $z=z_0(t)$ in Eq.
(\ref{Cartesianwarpmetric}). This reduces to $d\tau=d t$, taking into
account $d x=d y=0$ and $f(0)=1$.

\begin{figure}
\centering
\includegraphics[width=3.4in]{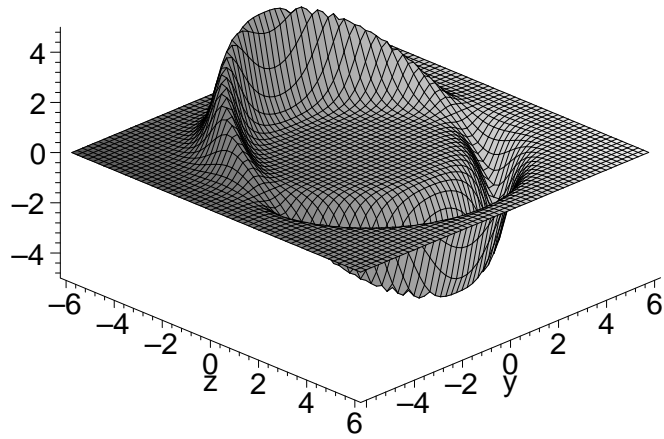}
\caption[The expansion of the volume elements for the Alcubierre
warp drive]{The plot depicts the expansion of the volume elements of an Alcubierre warp bubble moving along the positive $z$-axis, with an arbitrary velocity $v(t)$. Note that the volume elements are expanding behind the spaceship, and contracting in front of it. See the text for more datils.}\label{Alcubierre-expansion}
\end{figure}

Consider a spaceship placed within the Alcubierre warp bubble. The
expansion of the volume elements, $\theta=U^{\mu}{}_{;\mu}$, is
given by $\theta=v\;\left({\partial f}/{\partial z} \right)$.
Taking into account Eq. (\ref{E:form}), we have
\begin{equation}
\theta=v\;\frac{z-z_0}{r}\;\frac{d f(r)}{d r}.
\end{equation}
The center of the perturbation corresponds to the spaceship's
position $z_0(t)$. The volume elements are expanding behind the
spaceship, and contracting in front of it, as shown in Figure
\ref{Alcubierre-expansion}.

\subsection{Superluminal travel in the warp drive}

To demonstrate that it is possible to travel to a distant point and back in an arbitrary short time interval, consider two distant stars, $A$ and $B$, separated by a distance $D$ in flat spacetime. Suppose that, at the instant $t_0$, a spaceship moves away from $A$, using its engines, with a velocity $v<1$, and finally comes to rest at a distance $d$ from $A$. We shall, for simplicity, assume that $R\ll d\ll D$. Now, at this instant the perturbation of spacetime appears, centered around the spaceship's position, and pushing it away from $A$, and rapidly attains a constant acceleration, $a$. Consider that now at half-way between $A$ and $B$, the perturbation is modified, so that the acceleration rapidly varies from $a$ to $-a$. The spaceship finally comes to rest at a distance, $d$, from $B$, at which point the perturbation disappears. The spaceship then moves to $B$ at a constant velocity in flat spacetime. The return trip to $A$ is analogous.

Consider that the variations of the acceleration are extremely rapid, so that the
total coordinate time, $T$, in a one-way trip will be
\begin{equation}
T=2\left( \frac{d}{v}+\sqrt{\frac{D-2d}{a}} \right) \,,
\end{equation}
The proper time of an observer, in the exterior of the warp bubble, is equal to the coordinate time, as both are immersed in flat spacetime. The proper time
measured by observers within the spaceship is given by:
\begin{equation}
\tau=2\left( \frac{d}{\gamma v}+\sqrt{\frac{D-2d}{a}} \right)  \,,
\end{equation}
with $\gamma =(1-v^2)^{-1/2}$. The time dilation only appears in
the absence of the perturbation, in which the spaceship is moving
with a velocity $v$, using only it's engines in flat spacetime.

Using $R\ll d\ll D$, we can then obtain the following approximation $\tau\approx T\approx 2\sqrt{D/a}$, which can be made arbitrarily short, by increasing the value of $a$. This implies that the spaceship may travel faster than the speed of light, however, it moves along a spacetime temporal trajectory, contained within it's light cone, as light suffers the same distortion of spacetime \cite{Alcubierre}.

\subsection{The violation of the energy conditions}\label{warpENviolations}

\subsubsection{The violation of the WEC}

As mentioned in the previous chapters, the weak energy condition (WEC) states $T_{\mu\nu} \, U^{\mu} \, U^{\nu}\geq0$, in which $U^{\mu}$ is a timelike vector and $T_{\mu\nu}$ is the stress-energy tensor. As mentioned Chapters 9 and 10, 
its physical interpretation is that the
local energy density is positive, and by continuity it implies the
NEC. Now, one verifies that the WEC is violated for the warp drive metric, i.e.,
\begin{equation}
T_{\mu\nu} \; U^{\mu} \; U^{\nu}= -\frac{v^2}{32\pi}\; \left[
\left (\frac {\partial f}{\partial x} \right )^2 + \left (\frac
{\partial f}{\partial y} \right )^2 \right]  <0 \,,
\end{equation}
and by taking into account the Alcubierre form function
(\ref{E:form}), we have
\begin{equation}
T_{\mu\nu} \; U^{\mu} U^{\nu}= -\frac{1}{32\pi}\frac{v^2
(x^2+y^2)}{r^2} \left( \frac{d f}{d r} \right)^2<0
\label{WECviolation} \,.
\end{equation}

\begin{figure}
\centering
\includegraphics[width=3.8in]{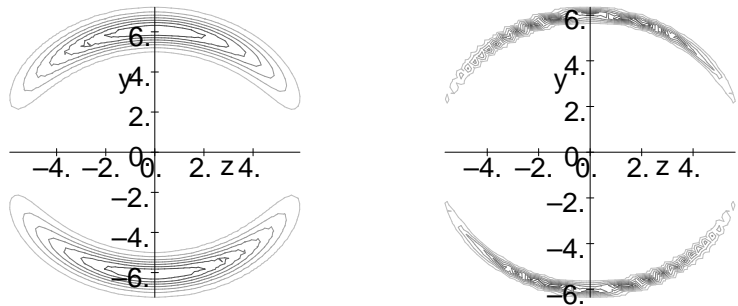}
\caption[Distribution of the energy density for the Alcubierre
warp drive]{The energy density is distributed in a toroidal region
perpendicular to the direction of travel of the spaceship, which
is situated at $z_0(t)$. We have considered the following values,
$v=2$ and $R=6$, with $\sigma=1$ and $\sigma=4$ in $(a)$ and
$(b)$, respectively. See the text for more details.}\label{Alcub-energydensity}
\end{figure}

By considering an orthonormal basis, we verify that the energy
density of the warp drive spacetime is given by
$T_{\hat{t}\hat{t}} = T_{\hat{\mu}\hat{\nu}} \; U^{\hat{\mu}} 
U^{\hat{\nu}}$, which is precisely given by Eq. (\ref{WECviolation}). It is easy to
verify that the energy density is distributed in a toroidal region
around the $z$-axis, in the direction of travel of the warp
bubble~\cite{PfenningF}, as may be verified from Figure
\ref{Alcub-energydensity}.  It is perhaps instructive to point out
that the energy density for this class of spacetimes is nowhere
positive\footnote{It is also interesting to note that the inclusion of a generic lapse function $\alpha(x,y,z,t)$, in the metric, decreases the negative energy density, which is now given by
\begin{equation}
T_{\hat{t}\hat{t}}= -\frac{v^2}{32\pi\,\alpha^2}\; \left[ \left
(\frac {\partial f}{\partial x} \right )^2 + \left (\frac
{\partial f}{\partial y} \right )^2 \right]  \,.
\end{equation}
One may impose that $\alpha$ may be taken as unity in the exterior and interior of the warp bubble, so that proper time equals coordinate time. In order to significantly decrease the negative energy density in the bubble walls, one may impose an extremely large value for the lapse function. However, the inclusion of the lapse function suffers from an extremely severe drawback, as proper time as measured in the bubble walls becomes absurdly large, $d\tau=\alpha\,dt$, for $\alpha \gg 1$.
}.

In analogy with the definitions in~\cite{visser2003,Kar2}, one may
quantify the ``total amount'' of energy condition violating matter
in the warp bubble, by defining the ``volume integral quantifier''\footnote{We refer the reader to  \cite{visser2003,Kar2} for details.}
\begin{eqnarray}
M_\mathrm{warp} = \int \rho_\mathrm{warp} \; d^3 x = \int
T_{\mu\nu} \; U^{\mu}\; U^{\nu}  \; d^3 x
        = -{v^2\over12} \int
\left( \frac{d f}{d r} \right)^2 \; r^2 \; d r.
\end{eqnarray}
This is not the total mass of the spacetime, but it
characterizes how much (negative) energy one needs to localize in
the walls of the warp bubble. For the specific shape function
(\ref{E:form}) we can estimate
\begin{equation}
M_\mathrm{warp} \approx - v^2 \; R^2 \; \sigma ,
\end{equation}
so that one can readily verify that the energy
requirements for the warp bubble scale quadratically with bubble
velocity and with bubble size, and inversely as the
thickness of the bubble wall \cite{LV-CQG}.

\subsubsection{The violation of the NEC}

As before mentioned in the previous chapters, the NEC states that $T_{\mu\nu} \, k^{\mu} \, k^{\nu}\geq0$, where
$k^{\mu}$ is \emph{any} arbitrary null vector and $T_{\mu\nu}$ is
the stress-energy tensor. The NEC for a null vector oriented along
the $\pm \hat z$ directions takes the following form
\begin{equation}
T_{\mu\nu} \; k^{\mu} \; k^{\nu}= -\frac{v^2}{8\pi}\, \left[ \left
(\frac{\partial f}{\partial x} \right )^2 + \left (\frac {\partial
f}{\partial y} \right )^2 \right] \pm \frac{v}{8\pi}\left( \frac
{\partial^{2}f}{\partial {x}^{2}} +
 \frac {\partial^{2}f}{\partial {y}^{2}}
\right)  \,.
\end{equation}
Note that if we average over the $\pm \hat z$ directions we have the following relation
\begin{equation}
{1\over2} \left[ T_{\mu\nu} \; k^{\mu}_{+\hat z} \;
k^{\nu}_{+\hat z} + T_{\mu\nu} \; k^{\mu}_{-\hat z} \;
k^{\nu}_{-\hat z} \right] = -\frac{v^2}{8\pi}\, \left[ \left
(\frac{\partial f}{\partial x} \right )^2 + \left (\frac {\partial
f}{\partial y} \right )^2 \right],
\end{equation}
which is manifestly negative, and implies that the NEC is violated for all
$v$. Furthermore, note that even if we do not average, the
coefficient of the term linear in $v$ must be nonzero
\emph{somewhere} in the spacetime. Then at low velocities this
term will dominate and at low velocities the un-averaged NEC will
be violated in either the $+\hat z$ or $-\hat z$ directions.

\subsection{The Quantum Inequality applied to the warp drive}

It is of interest to apply the QI to the warp drive spacetimes \cite{PfenningF}, and rather than deduce the QI in this section, we refer the reader to the Chapter on the Quantum Energy Inequalities. By inserting the energy density, Eq. (\ref{WECviolation}), into the QI, one arrives at the following inequality
\begin{equation}
t_0 \int_{-\infty}^{+\infty} {v(t)^2 \over r^2} \left({df\over
dr}\right)^2 {dt\over {t^2+t_0^2}} \leq {3\over \rho^2 \,t_0^4}
\,, \label{expr_QIwarp}
\end{equation}
where the quantity $\rho=\left(x^2+y^2 \right)^{1/2}$ is defined for notational
convenience.

One may consider that the warp bubble's velocity is roughly constant, i.e., $v_s(t) \approx v_b$, by assuming that the time scale of the sampling is sufficiently small compared to the time scale over which the bubble's velocity is varying. 
Now, taking into account the small sampling time, the $(t^2 +t_0^2)^{-1}$ term becomes strongly peaked, so that only a small portion of the geodesic is sampled by
the QI integral. Consider also that the observer is situated at the equator of
the warp bubble at $t=0$ \cite{PfenningF}, so that the geodesic is
approximated by $x(t) \approx f(\rho) v_b t $,
and consequently $r(t) = \left[ (v_b t)^2 (f(\rho) - 1)^2 + \rho^2
\right]^{1/2}$.

For simplicity, without a significant loss of generality, and instead of taking into account the Alcubierre form function (\ref{E:form}), one may consider a
piece-wise continuous form of the shape function given by
\begin{equation}\label{eq:pointwise}
f_{p.c.}(r) = 
\left\{
\begin{array}
{c}%
1 \qquad \qquad r < R-{\Delta\over2} \\
-{1\over\Delta}(r -R-{\Delta\over2})\qquad  R-{\Delta\over2}<r<
R+{\Delta\over2}\\
0 \qquad \qquad r > R+{\Delta\over2}
\end{array}
\right.  \,, 
\end{equation}
where $R$ is the radius of the bubble, and $\Delta$ the bubble
wall thickness \cite{PfenningF}. Note that $\Delta$ is related to the
Alcubierre parameter $\sigma$ by setting the slopes of the
functions $f(r)$ and $f_{p.c.}(r)$ to be equal at $r=R$, which
provides the following relationship $\Delta = \left[1+\tanh^2(\sigma R)\right]^2/[
 2\;\sigma\;\tanh(\sigma R)]$.

Note that in the limit of large $\sigma R$ one obtains the
approximation $\Delta \simeq 2/\sigma$, so that the QI-bound simplifies to
\begin{equation}
 t_0 \int_{-\infty}^{+\infty} {dt \over (t^2 + \bar{\beta}^2) (t^2+t_0^2)}
\leq {3 \Delta^2 \over v_b^2 \; t_0^4 \; \bar{\beta}^2} \,,
\end{equation}
where $\bar{\beta} = \rho/\left[ {v_b (1 - f(\rho) ) }\right]$. Now, evaluating the integral in the left-hand-side yields the following inequality
\begin{equation}
{\pi \over 3} \leq {\Delta^2 \over {v_b^2 \; t_0^4}} \left[{v_b
t_0 \over \rho} (1-f(\rho)) +1\right] \; . \label{wd-QI-bound}
\end{equation}

It is perhaps important to note that the above inequality is only
valid for sampling times on which the spacetime may be considered
approximately flat. 
Considering the Riemann tensor components in
an orthonormal frame \cite{PfenningF}, the largest component is
given by
\begin{equation}
|R_{{\hat t}{\hat y}{\hat t}{\hat y}}| = {3 v_b^2\; y^2 \over
4\;\rho^2}\left[ {d f(\rho) \over d\rho} \right]^2  \,,
\end{equation}
which yields $r_{\rm min} \equiv 1/ \sqrt{|R_{{\hat t}{\hat
y}{\hat t}{\hat y}}|} \sim 2\Delta /\left( {\sqrt 3}\; v_b \right)$, considering
$y= \rho$ and the piece-wise continuous form of the shape function. The sampling time must be smaller than this length scale, so that one may define $t_0=2 \alpha \,\Delta / \left( {\sqrt 3}\; v_b \right)$.
Assuming $\Delta / \rho \sim v_b\,t_0/\rho \ll 1$, the term
involving $1-f(\rho)$ in Eq. (\ref{wd-QI-bound}) may be neglected,
which provides
\begin{equation}
\Delta \leq {3\over 4}\sqrt{3\over\pi}\;{v_b \over \alpha^2}\, .
\end{equation}
Taking a specific value for $\alpha$, for instance, considering $\alpha = 1/10$, one obtains
\begin{equation}
\Delta \leq 10^2\, v_b\; L_{\rm Planck}\, ,
\label{eq:wall_thickness}
\end{equation}
where $L_{\rm Planck}$ is the Planck length.  Thus, unless $v_b$
is extremely large, the wall thickness cannot be much above the
Planck scale.

It is also interesting to find an estimate of the total amount of
negative energy that is necessary to maintain a warp metric. It
was found that the energy required for a warp bubble is on the
order of
\begin{equation}
E \leq - 3 \times 10^{20} \; M_{\rm galaxy} \; v_b\; ,
\end{equation}
which is an absurdly enormous amount of negative energy\footnote{Due to these results, one may tentatively conclude that the existence of these spacetimes is improbable. But, there are a series of considerations that can be applied to the QI. First, the QI is only of interest if one is relying on quantum field theory to provide the exotic matter to support Alcubierre warp bubble. However, there are classical systems (non-minimally coupled scalar fields) that violate the null and the weak energy conditions, whilst presenting plausible results when applying the QI (See Chapter 10). Secondly, even if one relies on quantum field theory to provide exotic matter, the QI does not rule out the existence of warp drive spacetimes, although they do place serious constraints on the geometry.}, roughly
ten orders of magnitude greater than the total mass of the entire
visible universe \cite{PfenningF}.

\section{Linearized warp drive}\label{ALsec:3}

In this section, we show that there are significant problems that arise even in the warp drive spacetime, even in weak-field regime, and long before strong field effects come into play. Indeed, to ever bring a warp drive into a strong-field regime, any highly-advanced civilization would first have to take it through the weak-field regime \cite{LV-CQG}. We now construct a more realistic model of a warp drive spacetime where the warp bubble interacts with a finite mass spaceship. To this effect, consider the linearized theory applied to warp drive spacetimes, for non-relativistic velocities, $v\ll 1$.


Consider now a spaceship in the interior of an Alcubierre warp
bubble, which is moving along the positive $z$ axis with a
non-relativistic constant velocity \cite{LV-CQG}, i.e., $v\ll 1$.
The metric is given by
\begin{eqnarray}
 d s^2&=&-d t^2+d x^2+d y^2+\left[d z-v\;f(x,y,z-vt)\,d t
\right]^2 
\nonumber   \\
&&-2\Phi(x,y,z-vt)\, \left[d t^2+d x^2+d y^2+(d
z-v\;f(x,y,z-vt)\,d t)^2\right] \label{warpspaceshipmetric}  \,,
\end{eqnarray}
where $\Phi$ is the gravitational field of the spaceship. If $\Phi =0$, the metric (\ref{warpspaceshipmetric}) reduces to the warp drive spacetime of Eq. (\ref{Cartesianwarpmetric}). If $v=0$, we have the metric representing the gravitational field of a static source.


We consider now the approximation by linearizing in the gravitational field of the
spaceship $\Phi$, but keeping the exact $v$ dependence .
For this case, the WEC is given by
\begin{eqnarray}
T_{\hat{\mu}\hat{\nu}} \; U^{\hat{\mu}}  U^{\hat{\nu}}=\rho -
\frac{v^2}{32\pi}\left[\left(\frac{\partial f}{\partial
x}\right)^2 + \left(\frac{\partial f}{\partial y}\right)^2\right]
+O(\Phi^2)
 \,,
\end{eqnarray}
and assuming the Alcubierre form function, we have
\begin{equation}
T_{\mu\nu} \; U^{\mu} U^{\nu}=\rho -\frac{1}{32\pi}\frac{v^2
(x^2+y^2)}{r^2} \left( \frac{d f}{d r} \right)^2 +O(\Phi^2)
 \,.
\end{equation}

Using the ``volume integral quantifier'', as before, we find the
following estimate
\begin{eqnarray}
\int T_{{\mu}{\nu}} \; U^{{\mu}}  U^{{\nu}} \;
d^3 x =  M_{\rm ship} -v^2 \;R^2\;\sigma + \int O(\Phi^2)  \;d^3
x\,,
\end{eqnarray}
which can be recast in the following form
\begin{equation}
M_\mathrm{ADM} = M_\mathrm{ship} + M_\mathrm{warp} + \int
O(\Phi^2)  \; d^3 x\,.
\end{equation}
Now, demanding that the volume integral of the WEC be positive, then we have
\begin{equation}
v^2 \;R^2\;\sigma  \leq M_{\rm ship} \,,
\end{equation}
which reflects the quite reasonable  condition that the net total energy stored in the warp field be less than the total mass-energy of the spaceship itself. Note that this inequality places a powerful constraint on the velocity of the warp bubble. 
Rewriting this constraint in terms of the size of the spaceship $R_\mathrm{ship}$ and the thickness of the warp bubble walls $\Delta = 1/\sigma$, we arrive at the following condition
\begin{equation}
v^2 \leq {M_{\rm ship}\over R_\mathrm{ship}}\; {R_\mathrm{ship} \;
\Delta\over R^2}.
\end{equation}
For any reasonable spaceship this gives extremely low bounds on
the warp bubble velocity.\\


One may analyse the NEC in a similar manner. Thus, the quantity $T_{{\mu}{\nu}} \, k^{{\mu}} k^{{\nu}}$ is given by
\begin{eqnarray}
T_{\mu\nu} \; k^{{\mu}} k^{{\nu}} &=& \rho
\pm \frac{v}{8\pi}\left(\frac{\partial^2 f}{\partial x^2} +
\frac{\partial^2 f}{\partial y^2}\right) -
\frac{v^2}{8\pi}\left[\left(\frac{\partial f}{\partial x}\right)^2
+ \left(\frac{\partial f}{\partial y} \right)^2\right] + O(\Phi^2)
\,.
\end{eqnarray}
Considering the ``volume integral quantifier'', we may estimate that
\begin{equation}
\int T_{\mu\nu} \; k^{{\mu}}  k^{{\nu}} \;
d^3 x = M_{\rm ship}  -v^2 \, R^2 \, \sigma   + \int O(\Phi^2) \;
d^3 x\,,
\end{equation}
which is [to order $O(\Phi^2)$] the same integral we encountered
when dealing with the WEC. 
In order to avoid that the total NEC violations in the warp field exceed the mass of the spaceship itself, we again demand that
\begin{equation}
v^2 \;R^2\;\sigma  \leq M_{\rm ship} \,,
\end{equation}
which places an extremely stringent condition on the linearized warp drive spacetime. More specifically, it reflects that that for all conceivably interesting situations the bubble velocity should be absurdly low, and it therefore appears unlikely that, by using this analysis, the warp drive will ever prove to be technologically useful. 

Finally, we emphasize that any attempt at building up a ``strong-field'' warp drive
starting from an approximately Minkowski spacetime will inevitably
have to pass through a weak-field regime. Taking into account the analysis presented above, we verify that the weak-field warp drives are already so tightly constrained, which implies additional difficulties for developing a ``strong field''
warp drive \footnote{See Ref. \cite{LV-CQG} for more details}.


\section{The horizon problem}\label{ALsec:4}

Shortly after the discovery of the Alcubierre warp drive solution \cite{Alcubierre}, an interesting feature of the spacetime was found. Namely, an observer on a spaceship cannot create nor control on demand an Alcubierre bubble, with $v>c$, around the ship. \cite{Krasnikov}. It is easy to understand this, by noting that an observer at the origin (with $t=0$), cannot alter events outside of his future light cone, $|r|\leq t$, with $r=(x^2+y^2+z^2)^{1/2}$. In fact, applied to the warp drive, it is a trivial matter to show that points on the outside front edge of the bubble are always spacelike separated from the centre of the bubble.

The analysis is simplified in the proper reference frame of an
observer at the centre of the bubble, so that using the transformation
$z'=z-z_{0}(t)$, the metric is given by
\begin{equation}
ds^2=-dt^2+dx^2+dy^2+\left[dz'+(1-f)vdt\right]^2  \,.
\end{equation}
Now, consider a photon emitted along the $+Oz$ axis (with
$ds^2=dx=dy=0$), so that the above metric provides $dz'/dt=1-(1-f)v $.
If the spaceship is at rest at the center of the bubble, then
initially the photon has $dz/dt = v + 1$ or $dz'/dt = 1$ (recall that
$f=1$ in the interior of the bubble). However, at a specific point
$z'=z'_c$, with $f=1-1/v$, we have $dz'/dt=0$ \cite{Everett}. Once
photons reach $z'_c$, they remain at rest relative to the bubble
and are simply carried along with it. This implies that photons emitted in the
forward direction by the spaceship never reach the outside edge of
the bubble wall, which therefore lies outside the forward light
cone of the spaceship. This behaviour is reminiscent of an {\it event horizon}. Thus, the bubble thus cannot be created, or controlled, by any action of the spaceship crew, which does not mean that Alcubierre bubbles, if it were possible to create them, could not be used as a means of superluminal travel. It only implies that the
actions required to change the metric and create the bubble must
be taken beforehand by some observer whose forward light cone
contains the entire trajectory of the bubble.\\


The appearance of an event horizon becomes evident in the 2-dimensional model of the Alcubierre space-time \cite{Hiscock,Clark,Gonz}. Consider that the axis of symmetry coincides with the line element of the spaceship, so that the metric (\ref{Cartesianwarpmetric}), reduces to
\begin{equation}
ds^2=-(1-v^2 f^2)dt^2-2vfdzdt+dz^2  \,.\label{2+1warpmetric}
\end{equation}
For simplicity, we consider a constant bubble velocity, $v(t)=v_{b}$, and $r=[(z-v_{b}t)^2]^{1/2}$. Note that the metric components of Eq. (\ref{2+1warpmetric}) only depend on $r$, which may be adopted as a coordinate, so if $z>v_{b}t$, we consider the transformation $r=(z-v_{b}t)$. Using this transformation, $dz=dr+v_{b}\,dt$, the metric (\ref{2+1warpmetric}) takes the following form
\begin{equation}
ds^2=-A(r)\left[dt-\frac{v_{b}(1-f(r))}{A(r)}\;dr\right]^2+\frac{dr^2}{A(r)}
\,,    \label{Hiscockwarp}
\end{equation}
where $A(r)$, denoted by the Hiscock function, is defined by $A(r)=1-v_{b}^2\,\left[1-f(r)\right]^2 $. 

Now, it is possible to represent the metric (\ref{Hiscockwarp}) in
a diagonal form, using a new time coordinate
\begin{equation}
d\tau=dt-\frac{v_{b}\,\left[1-f(r)\right]}{A(r)}\;dr  \,,
\end{equation}
so that the metric (\ref{Hiscockwarp}) reduces to a manifestly static form, given by
\begin{equation}
ds^2=-A(r)\,d\tau^2+\frac{dr^2}{A(r)}   \,.
\end{equation}

The $\tau$ coordinate has an immediate interpretation in terms of an observer
on board of a spaceship, namely, $\tau$ is the proper time of the
observer, taking into account that $A(r)\rightarrow 1$ in the limit $r\rightarrow
0$. We verify that the coordinate system is valid for any value of
$r$, if $v_{b}<1$. If $v_{b}>1$, we have a coordinate singularity
and an event horizon at the point $r_{0}$ in which
$f(r_{0})=1-1/v_b$ and $A(r_{0})=0$.

\section{Superluminal subway: The Krasnikov tube}\label{ALsec:5}

It was pointed out above, that an interesting aspect of the warp drive resides in the fact  that points on the outside front edge of a superluminal bubble are always spacelike separated from the centre of the bubble. This implies that an observer in a spaceship cannot create nor control on demand an Alcubierre bubble. However, causality considerations do not prevent the crew of a spaceship from arranging, by their own actions, to complete a round trip from the Earth to a distant star and back in an arbitrarily short time, as measured by clocks on the Earth, by altering the metric along the path of their outbound trip. Thus, Krasnikov introduced a metric with an interesting property that although the time for a one-way trip to a distant destination cannot be shortened \cite{Krasnikov}, the time for a round trip, as measured by clocks at the starting point (e.g. Earth), can be made arbitrarily short, as will be demonstrated below.

\subsection{The 2-dimensional Krasnikov solution}

The 2-dimensional Krasnikov metric is given by
\begin{eqnarray}
ds^2&=&-(dt-dx)(dt+k(t,x)dx)
     \nonumber   \\
&=&-dt^2+\left[1-k(x,t)\right]\,dx\,dt+k(x,t)\,dx^2 \,,
       \label{kras:2d-metric}
\end{eqnarray}
where the form function $k(x,t)$ is defined by
\begin{equation}
k(t,x)=1-(2-\delta)
\theta_{\varepsilon}(t-x)\left[\theta_{\varepsilon}(x)-
\theta_{\varepsilon}(x+\varepsilon-D)\right]   \,,
     \label{def:k}
\end{equation}
and $\delta$ and $\varepsilon$ are arbitrarily small positive
parameters. $\theta_{\varepsilon}$ denotes a smooth monotone
function
\[
\theta_{\varepsilon}(\xi)=\left\{ \begin{array}{ll}
                     1,   & {\rm if}\; \xi>\varepsilon \,, \\
                     0,    & {\rm if}\; \xi<0  \,.
                     \end{array}
                     \right.
\] \\

One may identify essentially three distinct regions in the Krasnikov two-dimensional
spacetime, which is summarized in the following manner.

\begin{description}

\item[{\bf The outer region:}]


The outer region is given by the following set
\begin{equation}
\{x<0\}\cup\{x>D\}\cup\{x>t\}   \,.
\end{equation}
The two time-independent $\theta_\epsilon$-functions between the square
brackets in Eq.~(\ref{def:k}) vanish for $x < 0$ and cancel for $x
> D$, ensuring $k=1$ for all $t$ except between $x=0$ and $x=D$.
When this behavior is combined with the effect of the factor
$\theta_\epsilon(t-x)$, one sees that the metric~(\ref{kras:2d-metric}) is flat, i.e., $k=1$, and reduces to Minkowski spacetime everywhere for $t<0$ and at all times outside the range $0<x<D$. Future light cones are generated by the following vectors: $r_{O}=\partial_{t}+\partial_{x}$ and $l_{O}=\partial_{t}-\partial_{x}$.\\

\item[{\bf The inner region:}]


The inner region is given by the following set
\begin{equation}
\{x<t-\varepsilon\}\cap\{\varepsilon<x<D-\varepsilon\}  \,,
\end{equation}
so that the first two $\theta_\epsilon$-functions in
Eq.~(\ref{def:k}) both equal $1$, while $\theta_\epsilon(x +
\epsilon - D) = 0$, giving $k = \delta -1$ everywhere within this
region. This region is also flat, but the light cones are {\it
more open}, being generated by the following vectors: $r_{I}=\partial_{t}+\partial_{x}$ and $l_{I}=-(1-\delta)\partial_{t}-\partial_{x}$.\\

\item[{\bf The transition region:}]


The transition region is a narrow curved strip in spacetime, with
width $\sim \varepsilon$. Two spatial boundaries exist between the
inner and outer regions. The first lies between $x=0$ and
$x=\varepsilon$, for $t>0$. The second lies between
$x=D-\varepsilon$ and $x=D$, for $t>D$. It is possible to view
this metric as being produced by the crew of a spaceship,
departing from point $A$ ($x=0$), at $t=0$, travelling along the
$x$-axis to point $B$ ($x=D$) at a speed, for simplicity,
infinitesimally close to the speed of light, therefore arriving at
$B$ with $t\approx D$.\\

\end{description}

\begin{figure}[h]
\centering
  \includegraphics[width=2.5in]{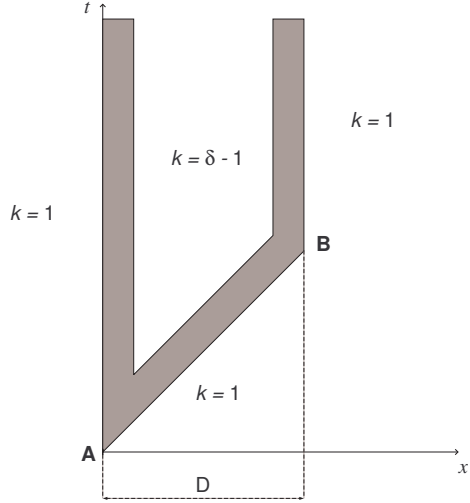}
  \caption[The Krasnikov spacetime in the $(x,t)$ plane]
  {The plot depicts the Krasnikov spacetime in the $(x,t)$ plane, where the vertical lines
  $A$ and $B$ are the world lines of the stars $A$ and $B$,
  respectively. The world line of the spaceship is approximately represented
  by the line segment $AB$. See the text for more details.}\label{fig:Kras-2d}
\end{figure}

Thus, the metric is modified by changing $k$ from $1$ to $\delta-1$
along the $x$-axis, in between $x=0$ and $x=D$, leaving a
transition region of width $\sim \varepsilon$ at each end for
continuity. However, as the boundary of the forward light cone of the
spaceship at $t=0$ is $|x|=t$, it is not possible for the crew to
modify the metric at an arbitrary point $x$ before $t=x$. This
fact accounts for the factor $\theta_{\varepsilon}(t-x)$ in the
metric, ensuring a transition region in time between the inner and
outer region, with a duration of $\sim \varepsilon$, lying along
the wordline of the spaceship, $x\approx t$. The geometry is shown
in the $(x,t)$ plane in Figure \ref{fig:Kras-2d}.


\subsection{Superluminal travel within the Krasnikov tube}

\begin{figure}[h]
\centering
  \includegraphics[width=4.0in]{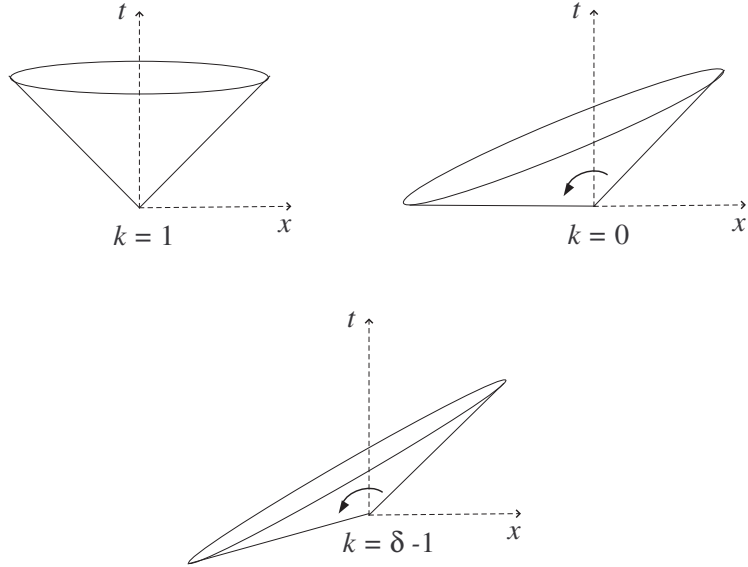}
  \caption[Forward light cones in the 2-dimensional Krasnikov
  spacetime]{Forward light cones in the 2-dimensional Krasnikov
  spacetime for $k=1$, $k=0$ and $k=\delta-1$.}\label{fig:Kras-lightcones}
\end{figure}

The factored form of the metric (\ref{kras:2d-metric}), for $ds^2=0$, provides some interesting properties of the spacetime with $\delta-1\leq k \leq 1$. Note that the two branches of the forward light cone in the $(t,x)$ plane are given
by $dx/dt=1$ and $dx/dt=-k$. As $k$ becomes smaller and then
negative, the slope of the left-hand branch of the light cone
becomes less negative and then changes sign. This implies that the light cone
along the negative $x$-axis opens out, as depicted in Figure
\ref{fig:Kras-lightcones}).

The inner region, with $k=\delta -1$, is flat because the metric (\ref{kras:2d-metric}) may be cast into the Minkowski form,
applying the following coordinate transformations
\begin{equation}
dt'=dt+\left( \frac{\delta}{2}-1 \right) dx\,,\qquad dx'=\left(
\frac{\delta}{2} \right) dx  \,,
     \label{Mink-tranf}
\end{equation}
and one verifies that the transformation is singular at $\delta=0$, i.e., $k=-1$. Note
that the left branch of the region is given by $dx'/dt'=-1$.

From the above analysis, one may easily deduce the following expression
\begin{equation}
\frac{dt}{dt'}=1+\left( \frac{2-\delta}{\delta}
\right)\frac{dx'}{dt'}  \,.
     \label{kras:time}
\end{equation}
For an observer moving along the positive $x'$ and $x$ directions,
with $dx'/dt'<1$, we have $dt'>0$ and consequently $dt>0$, if
$0<\delta \leq 2$. However, if the observer is moving sufficiently
close to the left branch of the light cone, given by $dx'/dt'=-1$,
Eq. (\ref{kras:time}) provides us with $dt/dt'<0$, for $\delta<
1$. Therefore we have $dt<0$, which means that the observer traverses backward in time, as measured by observers in the outer region, with $k=1$.\\

The superluminal travel analysis is as follows. Consider a spaceship departing from star $A$ and arriving at star $B$, at the instant $t\approx D$. Along this journey, the crew of the spaceship modify the metric, so that $k\approx -1$, for simplicity, along the trajectory. Now imagine that the spaceship returns to star $A$, travelling with a velocity arbitrarily close to the speed of light, i.e., $dx'/dt'\approx -1$. Therefore, from Eq. (\ref{Mink-tranf}), one obtains the following relation
\begin{equation}
v_{\rm return}=\frac{dx}{dt}\approx
-\frac{1}{k}=\frac{1}{1-\delta}\approx 1
\end{equation}
and $dt<0$, for $dx<0$. The return trip from star $B$ to $A$ is done
in an interval of $\Delta t_{\rm return}=-D/v_{\rm return}=D/(\delta
-1)$. Note that the total interval of time, measured at $A$, is given by $T_A
=D+\Delta t_{\rm return}=D \delta$. For simplicity, consider
$\varepsilon$ negligible, so that superluminal travel is implicit, as
$|\Delta t_{\rm return}|<D$, if $0<\delta < 1$, i.e.,
we have a spatial spacetime interval between $A$ and $B$. Now,
$T_A$ is always positive, but may attain a value arbitrarily close to
zero, for an appropriate choice of $\delta$.

Note that for the case $\delta<1$, it is always possible to choose
an allowed value of $dx'/dt'$ for which $dt/dt' = 0$, meaning that
the return trip is instantaneous as seen by observers in the
external region. This follows easily from Eq.~(\ref{kras:time}),
which implies that $dt/dt' = 0$ when $dx'/dt'$ satisfies $dx'/dt' = - \delta/(2- \delta)$, which lies between $0$ and $-1$ for $0< \delta < 1$.

\subsection{The 4-dimensional generalization}

Shortly after the Krasnikov two-dimensional solution, the analysis was generalized to four dimensions by Everett and Roman~\cite{Everett}, who denoted the solution as the {\it Krasnikov tube}. The latter four-dimensional modification of the metric begins along the path of the spaceship, which is moving along the $x$-axis, and occurs at the position $x$, at time $t \approx x$, which is the time of passage of the spaceship. Everett and Roman also assumed that the disturbance in the metric propagated radially outward from the $x$-axis, so that causality guarantees that at time $t$ the region in which the metric has been modified cannot extend beyond $\rho = t - x$, where $\rho={(y^2 + z^2)}^{1/2}$. The modification in the metric was also assumed to not extend beyond some maximum radial distance $\rho_{max} \ll D$ from the $x$-axis. 

Thus, the metric in the 4-dimensional
spacetime, written in cylindrical coordinates, is given by
\cite{Everett}
\begin{equation}
ds^2=-dt^2+[1-k(t,x,\rho)]\,dx \, dt+k(t,x,\rho)dx^2+d\rho^2+\rho^2
d\phi^2   \,,
   \label{4d-Krasnikov-metric}
\end{equation}
where the four-dimensional generalization of the Krasnikov form function is given by
\begin{equation}
k(t,x,\rho)=1-(2-\delta)\theta_{\varepsilon}(\rho_{max}-\rho)
\theta_{\varepsilon}(t-x-\rho)[\theta_{\varepsilon}(x)-
\theta_{\varepsilon}(x+\varepsilon-D)]   \,.
    \label{4d:form}
\end{equation}
For $t\gg D+\rho_{max}$ one has a tube of radius $\rho_{max}$
centered on the $x$-axis, within which the metric has been
modified. I is this structure that is denoted by the {\it Krasnikov tube}, and contrary to the Alcubierre spacetime metric, the metric of the Krasnikov tube is static, once it has been created.

The stress-energy tensor element $T_{tt}$ given by
\begin{equation}
T_{tt}=\frac{1}{32
\pi(1+k)^2}\left[-\frac{4(1+k)}{\rho}\frac{\partial k}{\partial
\rho}+3\left(\frac{\partial k}{\partial
\rho}\right)^2-4(1+k)\frac{\partial^2 k}{\partial \rho^2}\right]
\,,
\end{equation}
can be shown to be the energy density measured by a static
observer \cite{Everett}, and violates the WEC in a certain range
of $\rho$, i.e., $T_{\mu\nu}U^{\mu}U^{\nu}<0$. To this effect, consider the energy density in
the middle of the tube and at a time long after it's formation,
i.e., $x=D/2$ and $t\gg x+\rho +\varepsilon $, respectively. In
this region we have $\theta_{\varepsilon}(x)=1$,
$\theta_{\varepsilon}(x+\varepsilon-D)=0$ and
$\theta_{\varepsilon}(t-x-\rho)=1$. With this simplification the
form function (\ref{4d:form}) reduces to
\begin{equation}
k(t,x,\rho)=1-(2-\delta)\theta_{\varepsilon}(\rho_{max}-\rho)  \,.
     \label{4d:midtube-form}
\end{equation}
A useful form for $\theta_{\varepsilon}(\xi)$ \cite{Everett} is given by
\begin{equation}
\theta_{\varepsilon}(\xi)=\frac{1}{2}\left \{ \tanh \left
[2\left(\frac{2\xi}{\varepsilon}-1 \right )\right]+1 \right \} \,,
\end{equation}
so that the form function (\ref{4d:midtube-form}) yields
\begin{equation}
k=1-\left (1-\frac{\delta}{2}\right)\left \{ \tanh \left
[2\left(\frac{2\xi}{\varepsilon}-1 \right )\right]+1 \right \} \,.
\end{equation}

\begin{figure}
\centering
\includegraphics[width=7cm]{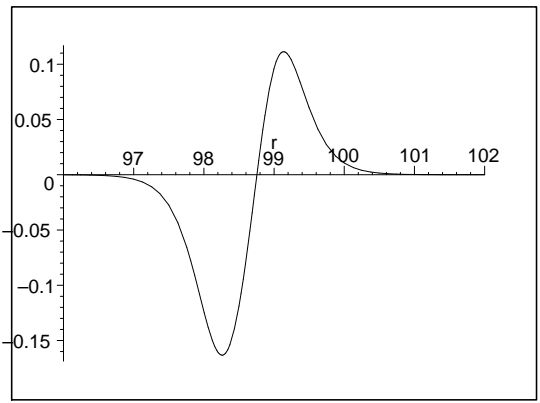}
\caption[Energy density of the Krasnikov tube]{Graph of the energy
density, $T_{tt}$, as a function of $\rho$ at the middle of the
Krasnikov tube, $x=D/2$, and long after it's formation, $t\gg
x+\rho +\varepsilon $. We consider the following values for the
parameters: $\delta =0.1$, $\varepsilon =1$ and
$\rho_{max}=100\varepsilon =100$. See the text for details.}
\label{fig3-superluminal}
\end{figure}

Choosing the following values for the parameters: $\delta =0.1$,
$\varepsilon =1$ and $\rho_{max}=100\varepsilon =100$, the
negative character of the energy density is manifest in the
immediate inner vicinity of the tube wall, as shown in Figure
\ref{fig3-superluminal}.

\section{Closed timelike curves}\label{ALsec:6}

\subsection{The warp drive}

Consider a hypothetical spaceship immersed within a warp bubble, moving along a timelike curve, with an arbitrary value of $v(t)$. Due to the latter, the metric of the warp drive permits superluminal travel, which raises the possibility of the existence of CTCs. Although the solution deduced by Alcubierre by itself does not possess CTCs, Everett demonstrated that these are created by a simple modification of the Alcubierre metric \cite{Everett}, by applying a similar analysis as is carried out using tachyons.

The modified metric takes the form
\begin{equation}
ds^2=-dt^2+dx^2+dy^2+(dz-vfdt)^2  \,,
   \label{modwarpmetric}
\end{equation}
with $v(t)=dz_0(t)/dt$ and $r(t)=[(z-z_{0})^2+(y-y_0)^2+z^2]^{1/2} $.
As in Section \ref{Alcubierrewarp} spacetime is flat in the exterior of a warp bubble with radius $R$, which now in the modified is centered in $(0,y_0,z_0(t))$. The bubble moves with a velocity $v$, on a trajectory parallel to the $z$-axis. Consider, for simplicity, the form function given by Eq. (\ref{E:form}). We shall also impose that $y_0\gg R$, so that the form function is negligible, i.e., $f(y_0)\approx 0$.

Now, consider two stars, $S_1$ and $S_2$, at rest in the
coordinate system of the metric (\ref{modwarpmetric}), and located
on the $z$-axis at $t=0$ and $t=D$, respectively. The metric along
the $z$-axis is Minkowskian as $y_0\gg R$. Therefore, a light beam
emitted at $S_1$, at $t=0$, moving along the $z$-axis with
$dz/dt=1$, arrives at $S_2$ at $t=D$. Suppose that the spaceship
initially starts off from $S_1$, with $v=0$, moving off to a
distance $y_0$ along the $y-$axis and neglecting the time it needs
to cover $y=0$ to $y=y_0$. At $y_0$, it is then subject to a
uniform acceleration, $a$, along the the $z-$axis for $0<z<D/2$,
and $-a$ for $D/2<z<D$. The spaceship will arrive at the spacetime
event $S_2$ with coordinates $z=D$ and $t=2\sqrt{D/a}\equiv T$.
Once again, the time required to travel from $y=y_0$ to $y=0$ is
negligible.

The separation between the two events, departure and arrival, is
$D^2-T^2=D^2(1-4/(aD)$ and will be spatial if $a > 4/D$ is verified.
In this case, the spaceship will arrive at $S_2$ before the light
beam, if the latter's trajectory is a straight line, and both
departures are simultaneous from $S_1$. Inertial observers
situated in the exterior of the spaceship, at $S_1$ and $S_2$,
will consider the spaceship's movement as superluminal, since the
distance $D$ is covered in an interval $T<D$. However, the
spaceship's wordline is contained within it's light cone. The
worldline of the spaceship is given by $z=vt$, while it's future
light cone is given by $z=(v\pm 1)t$. The latter relation can
easily be inferred from the null condition, $ds^2=0$.

Since the quadri-vector with components $(T,0,0,D)$ is spatial,
the temporal order of the events, departure and arrival, is not
well-defined. Introducing new coordinates, $(t',x',y',z')$,
obtained by a Lorentz transformation, with a boost $\beta$ along
the $z$-axis. The arrival at $S_2$ in the $(t',x',y',z')$
coordinates correspond to
\begin{equation}
T'=\gamma (2\sqrt{D/a}-\beta D)\,, \qquad Z'=\gamma
(D-2\sqrt{D/a})  \,,
\end{equation}
with $\gamma =(1-\beta ^2)^{-1/2}$. The events, departure and
arrival, will be simultaneous if $a=4/(\beta^2D)$. The arrival
will occur before the departure if $T'<0$, i.e.,
\begin{equation}
a> \frac{4}{\beta^2D}  \,.
    \label{spatialcharacter}
\end{equation}

The fact that the spaceship arrives at $S_2$ with $t'<0$, does not
by itself generate CTCs. Consider the metric (\ref{modwarpmetric}), substituting $z$ and $t$ by $\Delta
z'=z'-Z'$ and $\Delta t'=t'-T'$, respectively; $v(t)$ by $-v(t)$;
$a$ by $-a$; and $y_0$ by $-y_0$. This new metric describes a
spacetime in which an Alcubierre bubble is created at $t'=T'$,
which moves along $y=-y_0$ and $x=0$, from $S_1$ to $S_2$ with a
velocity $v'(t')$, and subject to an acceleration $a'$. For
observers at rest relatively to the coordinates $(t',x',y',z')$,
situated in the exterior of the second bubble, it is identical to
the bubble defined by the metric (\ref{modwarpmetric}), as it is
seen by inertial observers at rest at $S_1$ and $S_2$. The only
differences reside in a change of the origin, direction of
movement and possibly of the value of acceleration. The stars,
$S_1$ and $S_2$, are st rest in the coordinate system of the
metric (\ref{modwarpmetric}), and in movement along the
negative direction of the $z$-axis with velocity $\beta$,
relatively to the coordinates $(t',x',y',z')$. The two coordinate
systems are equivalent due to the Lorentz invariance, so if the
first is physically realizable, then so is the second. In the new
metric, by analogy with Eq. (\ref{modwarpmetric}), we have
$d\tau=dt'$, i.e., the proper time of the observer, on board of
the spaceship, travelling in the centre of the second bubble, is
equal to the time coordinate, $t'$. The spaceship will arrive at
$S_1$ in the temporal and spatial intervals given by $\Delta t'>0$
and $\Delta z'<0$, respectively. As in the analysis of the first
bubble, the separation between the departure, at $S_2$, and the
arrival $S_1$, will be spatial if the analogous relationship of
Eq. (\ref{spatialcharacter}) is verified. Therefore, the temporal
order between arrival and departure is also not well-defined. As
will be verified below, when $z$ and $z'$ decrease and $t'$
increases, $t$ will decrease and a spaceship will arrive at $S_1$
at $t<T$. In fact, one may prove that it may arrive at $t<0$.

Since the objective is to verify the appearance of CTCs, in
principle, one may proceed with some approximations. For
simplicity, consider that $a$ and $a'$, and consequently $v$ and
$v'$ are enormous, so that $T\ll D$ and $\Delta t'\ll -\Delta z'$.
In this limit, we obtain the approximation $T\approx 0$, i.e., the
journey of the first bubble from $S_1$ to $S_2$ is approximately
instantaneous. Consequently, taking into account the Lorentz
transformation, we have $Z'\approx \gamma D$ and $T'\approx
-\gamma \beta D$. To determine $T_1$, which corresponds to the
second bubble at $S_1$, consider the following considerations:
since the acceleration is enormous, we have $\Delta t'\approx 0$
and $\Delta t=T_1-T\approx T_1$, therefore $\Delta z=-D\approx
\gamma \Delta z'$ and $\Delta t\approx \gamma \beta \Delta z'$,
from which one concludes that
\begin{equation}
T_1\approx -\beta D<0  \,.
\end{equation}

\subsection{The Krasnikov tube}

As mentioned above, for superluminal speeds the warp drive metric has a horizon so that an observer in the center of the bubble is causally separated from the front edge of the bubble. Therefore he/she cannot control the Alcubierre bubble on demand. In order to address this problem, Krasnikov proposed a two-dimensional metric \cite{Krasnikov}, which was later extended to a four-dimensional model \cite{Everett}, as outlined in Section \ref{ALsec:5}. A two-dimensional Krasnikov tube does not generate CTCs. But the situation is quite different in the 4-dimensional generalization. Using two such tubes it is a simple matter, in principle, to generate CTCs \cite{Everett:1995nn}. The analysis is similar to that of the warp drive, so that it will be treated in summary.

Imagine a spaceship travelling along the $x$-axis, departing from
a star, $S_1$, at $t=0$, and arriving at a distant star, $S_2$, at
$t=D$. An observer on board of the spaceship constructs a
Krasnikov tube along the trajectory. It is possible for the
observer to return to $S_1$, travelling along a parallel line to
the $x$-axis, situated at a distance $\rho_0$, so that $D\gg
\rho_0\gg 2\rho_{max}$, in the exterior of the first tube. On the
return trip, the observer constructs a second tube, analogous to
the first, but in the opposite direction, i.e., the metric of the
second tube is obtained substituting $x$ and $t$, for $X=D-x$ and
$T=t-D$, respectively in eq. (\ref{4d-Krasnikov-metric}). The
fundamental point to note is that in three spatial dimensions it
is possible to construct a system of two non-overlapping tube
separated by a distance $\rho_0$.

After the construction of the system, an observer may initiate a
journey, departing from $S_1$, at $x=0$ and $t=2D$. One is only
interested in the appearance of CTCs in principle, therefore the
following simplifications are imposed: $\delta$ and $\varepsilon$
are infinitesimal, and the time to travel between the tubes is
negligible. For simplicity, consider the velocity of propagation
close to that of light speed. Using the second tube, arriving at
$S_2$ at $x=D$ and $t=D$, then travelling through the first tube,
the observer arrives at $S_1$ at $t=0$. The spaceship has
completed a CTC, arriving at $S_1$ before it's departure.

\section{Summary and Conclusion}\label{ALsec:conclusion}

In this chapter, we have seen how warp drive spacetimes can be used as gedanken-experiments to probe the foundations of general relativity. Though they are useful toy models for theoretical investigations, we emphasize that as potential technology they are greatly lacking. We have verified that exact solutions of the warp drive spacetimes necessarily violate the classical energy conditions, and continue to do so for arbitrarily low warp bubble velocity. Thus, the energy condition violations in this class of spacetimes is generic to the form of the geometry under consideration and is not simply a side-effect of the superluminal  properties. Furthermore, by taking into account the notion of the ``volume integral quantifier'', we have also verified that the ``total amount'' of energy condition violating matter in the warp bubble is negative. 

Using linearized theory, a more realistic model of the warp drive spacetime was constructed where the warp bubble interacts with a finite mass spaceship. The energy conditions were determined to first and second order of the warp bubble velocity, which safely ignores the causality problems associated with ``superluminal''
motion. A fascinating feature of these solutions resides in the fact that such a spacetime appear to be examples of a ``reactionless'' drives, where the warp bubble moves by interacting with the geometry of spacetime instead of expending reaction mass, and the spaceship is simply carried along with it. Note that in linearized theory the spaceship can be treated as a finite mass object placed within the warp bubble. It was verified that in this case, the ``total amount'' of energy condition violating matter, the ``net'' negative energy of the warp field, must be an appreciable fraction of the positive mass of the spaceship carried along by the warp bubble. This places an extremely stringent condition on the warp drive spacetime, in that the bubble velocity should be absurdly low. Finally, we point out that any attempt at building up a ``strong-field'' warp drive starting from an approximately Minkowski spacetime will inevitably have to pass through a weak-field regime. Since the weak-field warp drives are already so tightly constrained, the analysis of the linearized warp drive implies additional difficulties for developing a ``strong field'' warp drive.

Furthermore, we have shown that shortly after the discovery of the Alcubierre warp drive solution it was found that an observer on a spaceship cannot create nor control on demand a superluminal Alcubierre bubble, due to a feature that is reminiscent of an  event horizon. Thus, the bubble cannot be created, nor controlled, by any action of the spaceship crew. We emphasize that this does not mean that Alcubierre bubbles could not be theoretically used as a means of superluminal travel, but that the actions required to change the metric and create the bubble must be taken beforehand by an observer whose forward light cone contains the entire trajectory of the bubble.
To contour this difficulty, Krasnikov introduced a two-dimensional metric in which the time for a round trip, as measured by clocks at the starting point (e.g. Earth), can be made arbitrarily short. This metric was generalized the analysis to four dimensions, denoting the solution as the {\it Krasnikov tube}. It was also shown that this solution violates the energy conditions is specific regions of spacetime. Finally, it was shown that these spacetimes induce closed timelike curves.

\section*{Acknowledgements}

FSNL acknowledges financial support of the Funda\c{c}\~{a}o para a Ci\^{e}ncia e Tecnologia through an Investigador FCT Research contract, with reference IF/00859/2012, funded by FCT/MCTES (Portugal).



\begin{thebibliography}{100}


\bibitem{Morris}
M. S. Morris and K. S. Thorne, ``Wormholes in spacetime and their
use for interstellar travel: A tool for teaching General
Relativity,'' Am. J. Phys. {\bf 56}, 395 (1988).

\bibitem{VisserAL}
M. Visser, {\it Lorentzian Wormholes: From Einstein to Hawking}
(American Institute of Physics, New York, 1995).

\bibitem{Alcubierre}
M.~Alcubierre, ``The warp drive: hyper-fast travel within general
relativity,'' Class.\ Quant.\ Grav. {\bf 11}, L73-L77 (1994).

\bibitem{VB}
M.~Visser, B.~Bassett and S.~Liberati, ``Perturbative superluminal
censorship and the null energy condition,'' Proceedings of the
Eighth Canadian Conference on General Relativity and Relativistic
Astrophysics (AIP Press) (1999). 


\bibitem{VBL}
M.~Visser, B.~Bassett and S.~Liberati, ``Superluminal
censorship,'' Nucl.\ Phys.\ Proc.\ Suppl. {\bf 88}, 267-270 (2000).


\bibitem{Olum}
K.~Olum, ``Superluminal travel requires negative energy density,''
Phys.\ Rev.\ Lett, {\bf 81}, 3567-3570 (1998).


\bibitem{VisserEC}
C.~Barcelo and M.~Visser, ``Twilight for the energy conditions?,''
Int. J. Mod. Phys. D {\bf 11}, 1553 (2002). 



\bibitem{barcelovisser1}
C.~Barcelo and M.~Visser, ``Scalar fields, energy conditions, and
traversable wormholes,'' Class.\ Quant.\ Grav.\ {\bf 17}, 3843
(2000). 



\bibitem{barcelovisserPLB99}
C.~Barcelo and M.~Visser, ``Traversable wormholes from massless
conformally coupled scalar fields,'' Phys.\ Lett.\ B {\bf 466}, 127 (1999). 

\bibitem{Riess}
A.~G.~Riess {\it et al.}, ``Type Ia Supernova Discoveries at $z>1$
From the Hubble Space Telescope: Evidence for Past Deceleration
and Constraints on Dark Energy Evolution,'' Astrophys. J. {\bf
607}, 665-687 (2004). 


\bibitem{jerk}
M.~Visser, ``Jerk, snap, and the cosmological equation of state,''
Class.\ Quant.\ Grav.\  {\bf 21}, 2603 (2004).

\bibitem{rip}
R.~R.~Caldwell, M.~Kamionkowski and N.~N.~Weinberg, ``Phantom
Energy and Cosmic Doomsday,'' Phys.\ Rev.\ Lett.\  {\bf 91} 071301
(2003). 


\bibitem{Krasnikov}
S.~V.~Krasnikov, ``Hyper-fast Interstellar Travel in General
Relativity,'' Phys.\ Rev.\ D  {\bf 57}, 4760 (1998)
[arXiv:gr-qc/9511068].

\bibitem{Everett}
A.~E.~Everett and T.~A.~Roman, ``A Superluminal Subway: The
Krasnikov Tube,'' Phys. Rev. D {\bf 56}, 2100 (1997).

\bibitem{Natario}
J.~Nat\'{a}rio, ``Warp drive with zero expansion,'' Class.\
Quant.\ Grav. {\bf 19}, 1157, (2002). 

\bibitem{Ford:1994bjAL} 
  L.~H.~Ford and T.~A.~Roman,
  ``Averaged energy conditions and quantum inequalities,''
  Phys.\ Rev.\ D {\bf 51}, 4277 (1995).

\bibitem{Ford:1995wg} 
  L.~H.~Ford and T.~A.~Roman,
  ``Quantum field theory constrains traversable wormhole geometries,''
  Phys.\ Rev.\ D {\bf 53}, 5496 (1996).

\bibitem{PfenningF}
M.~J.~Pfenning and L.~H.~Ford, ``The unphysical nature of warp
drive,'' Class.\ Quant.\ Grav.\ {\bf 14}, 1743, (1997).

\bibitem{VanDenBroeck:1999sn} 
  C.~Van Den Broeck,
  ``A 'Warp drive' with reasonable total energy requirements,''
  Class.\ Quant.\ Grav.\  {\bf 16}, 3973 (1999).

\bibitem{GravelPlante}
P.~Gravel and J.~Plante, ``Simple and double walled Krasnikov
tubes: I. Tubes with low masses,'' Class.\ Quant.\ Grav.\ {\bf
21}, L7, (2004).

\bibitem{Gravel}
P.~Gravel, ``Simple and double walled Krasnikov tubes: II.
Primordial microtubes and homogenization,'' Class.\ Quant.\ Grav.\
{\bf 21}, 767, (2004).

\bibitem{LV-CQG}
F.~S.~N.~Lobo and M.~Visser, ``Fundamental limitations on `warp
drive' spacetimes,'' Class.\ Quant.\ Grav.\ {\bf 21}, 5871 (2004).

\bibitem{York79} J. W. York, ``Kinematic and dynamics of general
  relativity'', in L. L. Smarr editor, {\it Sources of gravitational
  radiation''}, Cambridge University Press, UK, 1979, pp. 83-126.

\bibitem{Alcubierre08} M. Alcubierre, {\it Introduction to 3+1 numerical
relativity} (Oxford University Press, UK, 2008).

\bibitem{visser2003}
M. Visser, S. Kar and N. Dadhich, ``Traversable wormholes with
arbitrarily small energy condition violations,'' Phys. Rev. Lett.
{\bf 90}, 201102 (2003). 

\bibitem{Kar2}
S.~Kar, N.~Dadhich and M.~Visser, ``Quantifying energy condition
violations in traversable wormholes,'' Pramana {\bf 63}, 859-864
(2004). 

\bibitem{Hiscock}
W. A. Hiscock, ``Quantum effects in the Alcubierre warp drive
spacetime,'' Class. Quant. Grav. {\bf 14}, L183 (1997).

\bibitem{Clark}
C.~Clark, W.~A. Hiscock and S.~L. Larson, ``Null geodesics in the
Alcubierre warp drive spacetime: the view from the bridge,''
Class. Quant. Grav. {\bf 16}, 3965 (1999). 

\bibitem{Gonz}
P.~F. Gonz\'{a}lez-D\'{\i}az, ``On the warp drive space-time,''
Phys. Rev. D {\bf 62}, 044005 (2000). 

\bibitem{Everett:1995nn} 
  A.~E.~Everett,
  ``Warp drive and causality,''
  Phys.\ Rev.\ D {\bf 53}, 7365 (1996).


\end{thebibliography}
\end{document}